\documentclass[11pt]{article}
\usepackage{hyperref}
\pdfoutput=1
\begin{document}
\title{Aerodynamics of the Smallest Flying Insects}
\author{Laura A. Miller$^\dagger$, Steven Harenber$^\dagger$, Ty Hedrick$^\ddagger$, Alice Robinson$^\ddagger$,\\
 Arvind Santhanakrishnan$^\star$, and Audrey Lowe$^\dagger$ \\
\\\ $\dagger$ Department of Mathematics\\
$\ddagger$ Department of Biology\\
University of North Carolina, Chapel Hill, NC 27599, USA\\
$\star$ Department of Biomedical Engineering\\
Georgia Institute of Technology, Atlanta, GA 30332, USA}
\maketitle
\begin{abstract}
We present fluid dynamics videos of the flight of some of the smallest insects including the jewel wasp, \textit{Ampulex compressa}, and thrips, \textit{Thysanoptera} spp. The fruit fly, \textit{Drosophila melanogaster}, is large in comparison to these insects. While the fruit fly flies at $Re \approx 120$, the jewel wasp flies at $Re \approx 60$, and thrips flies at $Re \approx 10$. Differences in the general structures of the wakes generated by each species are observed. The differences in the wakes correspond to changes in the ratio of lift forces (vertical component) to drag forces (horizontal component) generated. \href{http://manowar.amath.unc.edu/~lam9/movies/Miller_largeaps_movie.m4v}{Large} and 
\href{http://manowar.amath.unc.edu/~lam9/movies/Miller_smallaps_movie.mp4}{small} versions of the movies may be found by clicking the links.

\end{abstract}
\section{Introduction}

The fruit fly is considered to be one of the smallest flying insects with a wingspan of about 6 mm. It flies at about 200 wingbeats per second, which yields a Reynolds number of about 120. However, there is a large variety of flying insects that are much smaller in size. One group includes parasatoid wasps. The jewel wasp is one example and has a wingspan of about 3 mm, flying at a Reynolds number of about 60. Another group includes \textit{Thysanoptera}, commonly known as thrips. These insects are characterized by bristled wings and fly at Reynolds numbers around 10.

In 'typical' insect flight, lift is produced when a leading edge vortex is formed and remains attached to the wing, and a trailing edge vortex is formed and separates from the wing during each stroke. For tiny insect flight, neither the leading nor trailing edge vortices separate from the wing during the duration of each stroke.  As the Reynolds number is lowered from 100 to 1, the relative lift forces produced during flapping flight decrease while drag forces increase significantly. These transitions are associated with a change in the behavior of the vortex wakes behind the flapping wings~\cite{Miller:04}.

\section{Methods}
Thrips were collected during July and August locally on the University of North Carolina campus. Fruit flies were obtained from a local collection in the UNC Department of Biology. Jewel wasps were ordered from Carolina Biological. High speed videos of the insects were taken at 4000 Hz. In each case, one or two cameras were focused on a standard pipette tip from which the insects were filmed during takeoff. 

Immersed boundary simulations~\cite{Peskin} were performed for a two-dimensional wing with a fixed angle of attack of $45\deg$ immersed in a channel. Reynolds number was calculated as $\frac{\rho UL}{\mu}$, where $U$ was the maximum velocity in the channel, $\mu$ was the dynamic viscosity of the fluid, $\rho$ was the density of the fluid, and $L$ was the chord length of the wing. The initially velocity of the channel was set to zero and then linearly increased to a set velocity determined by the desired Reynolds number. 

Particle image velocimetry was used to quantify the flow fields around a dynamically scaled physical model of a wing that rotated about its base. The velocities of particles illuminated in the laser sheet were determined from sequential images analyzed using a cross-correlation algorithm (LaVision Inc. software). Image
pairs were analyzed with shifting overlapping interrogation windows of decreasing size ($64 \times 64$
pixels then $32 \times 32$ pixels). Lift and drag forces were also measured using a dynamically scaled physical model that rotated about its base~\cite{Sane}.

\end{document}